# FRACTAL DIMENSION OF STAR CLUSTERS

Ussipov N. [1], Akhmetali A.[1]*, Zaidyn M.[1], Akniyazova A.[1], Sakan A.[1], Kalambay M.[1,2,3,4,5], Shukirgaliyev B.[2,3,4,5]

[1] Department of Solid State Physics and Nonlinear Physics, Al-Farabi Kazakh National University, Almaty, Kazakhstan,
[2] Energetic Cosmos Laboratory, Nazarbayev University, Astana, Kazakhstan
[3] Fesenkov Astrophysical Institute, Almaty, Kazakhstan
[4] Heriot-Watt University, Edinburgh, UK
[5] Energetic Cosmos Laboratory, Nazarbayev University, Astana, Kazakhstan
*Corresponding author: akhmetalialmat@gmail.com

**Abstract.** *Quantitative analysis of the structure of star clusters is crucial for understanding their formation and evolution. In this article, we explore the application of fractal dimension analysis to study the evolution of star clusters, also fractal dimension, a concept from fractal geometry, provides a quantitative measure of the complexity and self-similarity of geometric objects. By considering star clusters as complex networks, we employ the box covering method to calculate their fractal dimension. Our methodology combines the well-established Minimum Spanning Tree (MST) and Box-Covering (BC) methods and using these methods, the fractal structure of the clusters was determined. It was revealed that star clusters disintegrate at a fractal dimension of 1.3 and obey a power law. It should be noted that the obtained result was compared with the results of the McLuster.*

## 1. Introduction

A star cluster is a gravitationally bound complex stellar structure with a radius ranging from 0.5 to several parsecs and a mass ranging from $10^3$ to $10^7$ solar masses [1].

Clusters of stars originate within massive molecular clouds triggered by the gravitational collapse of dense gas concentrations [2-3]. The influential impact of massive stars comes from various forms of stellar feedback, such as ionizing radiation, stellar winds, and radiation pressure, which can swiftly dismantle an entire molecular cloud [4-5]. The ejection of gas occurs at an approximate velocity of 10 km/s on average [6, 7]. Consequently, for star-forming areas with dimensions smaller than 10 pc, gas could be blown within a span of less than 1 million years. The star formation process is assessed through a quantitative measure known as the star formation efficiency (SFE), revealing the connection between neutral gas and the law of star formation [8, 9]. Star Formation Efficiency (SFE) is measured when a gas turns into stars. In nearby areas where star formation takes place, the SFE is estimated to be less than 30% [2, 10-11]. On a larger galactic scale, the average SFE is only a few percent [12]. When gas is ejected quickly and star formation efficiency is low, it can destroy star clusters [13-14]. If a cluster experiences gas expulsion, it will lose both mass and density [14-17]. Star clusters are direct products of the star formation process within galaxies and exhibit a significantly brighter nature compared to individual stars. Consequently, star clusters represent essential objects of study in star formation research [18].

Most star clusters have a hierarchical structure [19-21]. Quantitative and objective statistical measurements are used to comprehend the evolution of complex hierarchical structures of star clusters [22]. Additionally, it is important to evaluate the observed clusters using numerical modeling methods. This paper is aimed at studying the structure of star clusters through the analysis of their fractal dimension, considering star clusters as a complex network [23]. The primary issue in analyzing the fractality of complex networks lies in selecting a method for determining the dimension. When assessing the fractal properties of complex networks, the box-covering method is commonly employed [24]. The goal of this method is to identify the minimum number of boxes required to cover the entire network. Numerous algorithms have been proposed to address this challenge [25]. Minimum Spanning Tree (MST) was initially introduced in astronomy by Barrow et al. (1985) and finds applications in cosmology for the classification of large-scale structures, as well as for dynamic mass segregation in star clusters [26-31].

In this study, we introduce a novel approach by integrating Box-Covering (BC) and MST methods to analyze the fractal properties of star clusters across different stages of their evolution. While both methods have been applied separately in the study of fractal systems, their combination in the context of stellar clusters is innovative. The BC method is adapted to account for stellar mass loss during evolutionary stages, correlating this mass change with shifts in the cluster's fractal structure. Additionally, MST, which has traditionally been used to describe static topological structures, is applied dynamically to track the fragmentation and collapse of star-forming regions over time. This dual application of BC and MST provides new insights into the evolving fractal properties of star clusters, offering a deeper understanding of their structural development throughout their lifecycle.

In Section 2, we present a proposed methodology that explains the process of analyzing the fractal dimensions of star clusters with varying SFE, considering them as complex networks. In Section 3, we present and discuss our results, while the main conclusions of our work are outlined in Section 4.

## 2. Methodology

In this study, we calculated the fractal dimensions of star clusters at various stages of evolution. We ran simulations of SCs with different SFE, including 0.15, 0.17 and 0.20. The initial mass of the cluster was $M = 6000$ solar masses and it was simulated on a solar orbit, 8178 pc away from the Galactic center. The density profile of the initial cluster was modeled on the Plummer model [32]. After the start of the simulation, the effect of the gas explosion and the gravity of the host galaxy greatly change the shape of the initial cluster (see APPENDIX Fig. A1). And the lifetime of clusters can vary based on the SFE, so SCs with higher SFEs live longer [16-17] (also see the Table 1 in APPENDIX 2). We have created a star cluster model with a coefficient of (lambda) $\lambda = 0.05$ (1).

$$\lambda = \frac{r_h}{r_j} \quad (1)$$

The Jacobi radius, $r_j$ refers to the distance from the center of a cluster at which its gravitational potential becomes stronger than that of the host galaxy. And, the half-mass radius, $r_h$ is the radius within which half of the cluster's total mass is located [12]. Since many stars fly apart into the galaxy's space, the cluster loses its stars, reducing the $r_j$ and the $r_h$ (red circle in APPENDIX 1 Fig. A1).

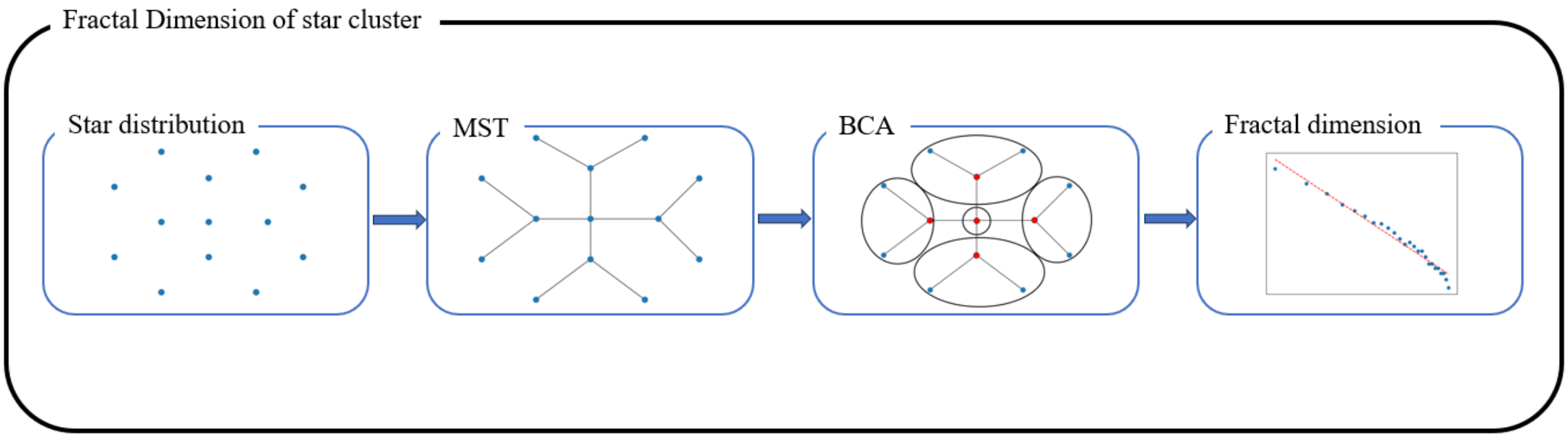

**Fig.1.** The flowchart demonstrates the algorithm of this study

.

To quantify the fractal dimension of the simulated star clusters, we used the MST and BC methods.

The Minimum Spanning Tree (MST) is a unique network with minimal weight that connects any distribution of points in space. In this context, the weight of the tree is understood as the sum of the weights of the edges included in the network. However, in our case, instead of considering the weights of the edges, we evaluate the distances of stars relative to each other. The MST algorithm should build a network by connecting stars within a cluster with a minimum total length. The resulting MST represents the basis of the cluster's connectivity, revealing the underlying hierarchical organization and relationships between the stars The minimum spanning tree can be used to determine the network topology [32-33]. The next step was to determine the fractal dimension of the network using the well-known box coverage method [26]. The box coverage method is widely used to calculate the fractal dimension of complex networks, but its effectiveness relies on determining the minimum number of boxes needed to cover the entire network. If the distribution of the number of boxes, $N_b(l_b)$, corresponds to a power law, it indicates that $D$ represents the fractal dimension of a complex network. The concept of fractal dimension, as a measure of occupancy and complexity within object space, holds great importance for the study of the reliability of complex networks [33-34].

$$N_b(l_b) \sim l_b^{-D} \tag{2}$$

where $l_b$ is the radius of the boxes, $N_b$ is the number of boxes.

The discovery of fractal features and self-similarity in star clusters not only opens up a new perspective for improving our understanding of the internal structure and characteristics of these clusters but also provides a new basis for explaining the mechanisms underlying their formation, evolutionary processes, and the coexistence of various characteristics within them.

## 3. Results and discussion

According to the calculations, it can be asserted that the star cluster has a certain fractal dimension. As an example, Figure 2 illustrates the results of calculating the fractal dimension for SFE=0.15 at the age of 10 million years. It is evident that the number of boxes decreases rapidly, depending on the radii, and follows a power law. However, there is a noticeable deviation from the power-law behavior when $\log(l_b) > 3.5$. This flattening suggests that at larger box sizes, the fractal nature of the star cluster begins to break down. In this case, the deviation is likely due to the fact that, at larger scales, the global structure of the cluster becomes more coherent and less fragmented, resulting in fewer boxes required to cover the structure.

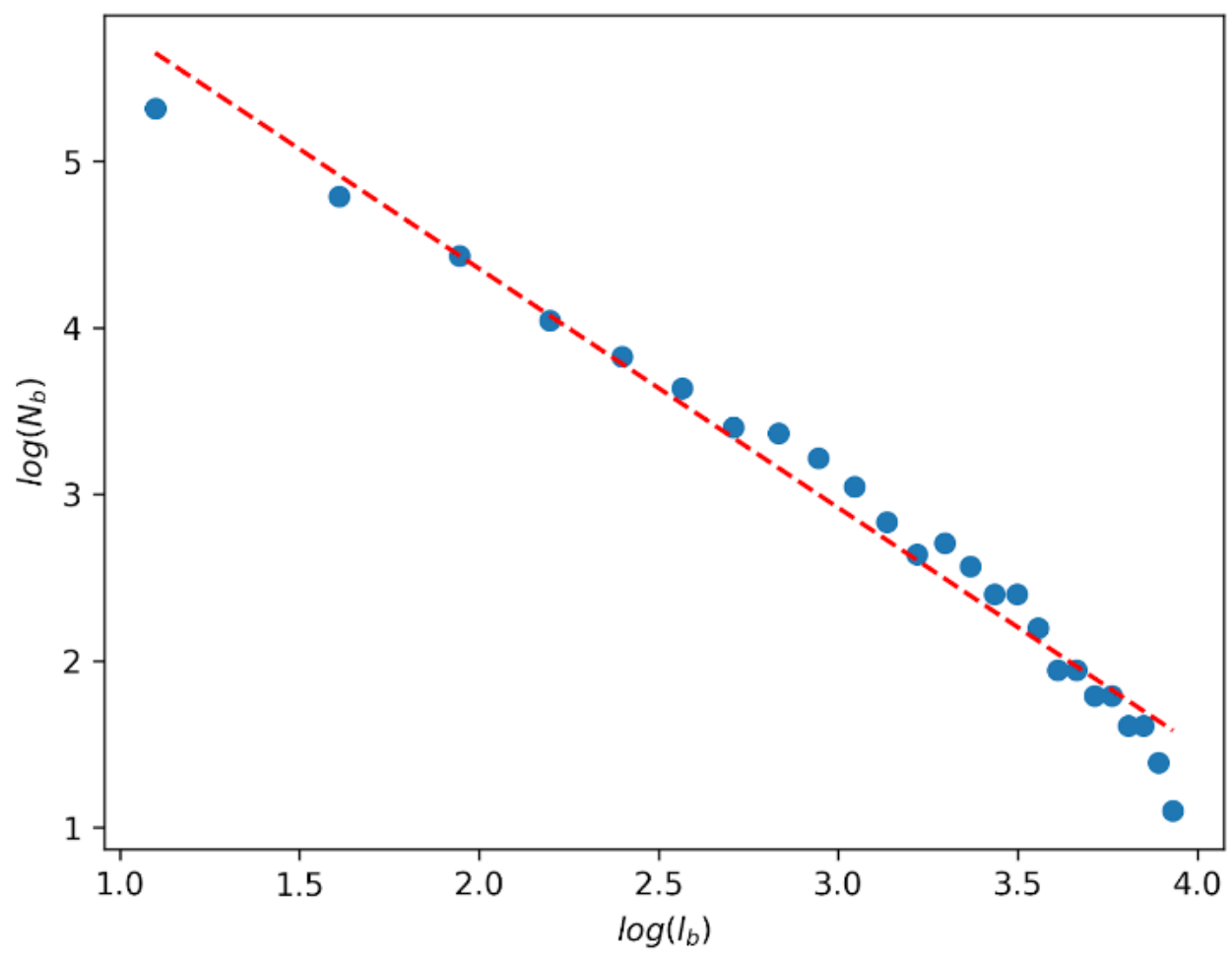

**Fig.2.** Fractal dimension of the cluster with SFE = 0.15 at 10 Myr age.

The main results of our study are illustrated in Figure 3. We compared the fractal dimensions for various SFE values in the time range of 50-950 million years for a Jacobi radius of 1. At SFE=0.15, the fractal dimension values exhibit a sharp decrease over time, while at SFE=0.17 and SFE=0.20, the rates of change are the same. It is evident that the fractal dimension of the star cluster depends on its evolution and undergoes changes with alterations in its structure.

Figure 3 shows the dynamics of changes in fractal dimension during the evolution of star clusters. In the initial stages of young cluster evolution, a consistent increase in fractal dimension is observed, reaching its peak value after approximately 200 million years. This phenomenon can be explained by a more complex structure and active cluster formation processes that enhance fractal characteristics. The growth of fractal dimension observed at later times in Figure 3 reflects the re-structuring of the cluster as new hierarchical patterns form due to dynamical interactions and external perturbations. As the cluster evolves, mass loss from stellar evolution (particularly from massive stars) and gravitational interactions between stars lead to the emergence of new substructures, increasing the fractal dimension. The increase in fractal dimension at 800 million years, for instance, is associated with the formation of subclusters or the merging of previously distinct structures.

However, after reaching the specified time interval, it becomes evident that the clusters start to lose their fractal dimension. This phenomenon is attributed to the loss of complexity or organization within the cluster's structure. It can be triggered by dynamic interactions between stars, leading to a gradual reduction in the number of stars in the cluster and, ultimately, a decrease in its fractal dimension. In the final stage of their life, clusters lose their structure and become more dispersed [9]. From Figure 3, it can be noted that when the fractal dimension of a cluster falls below a certain 'threshold' value of 1.3, the cluster is prone to decay.

Thus, analyzing the dynamics of changes in fractal dimension during the evolution of star clusters allows us to track the processes that shape and dismantle cluster structures.

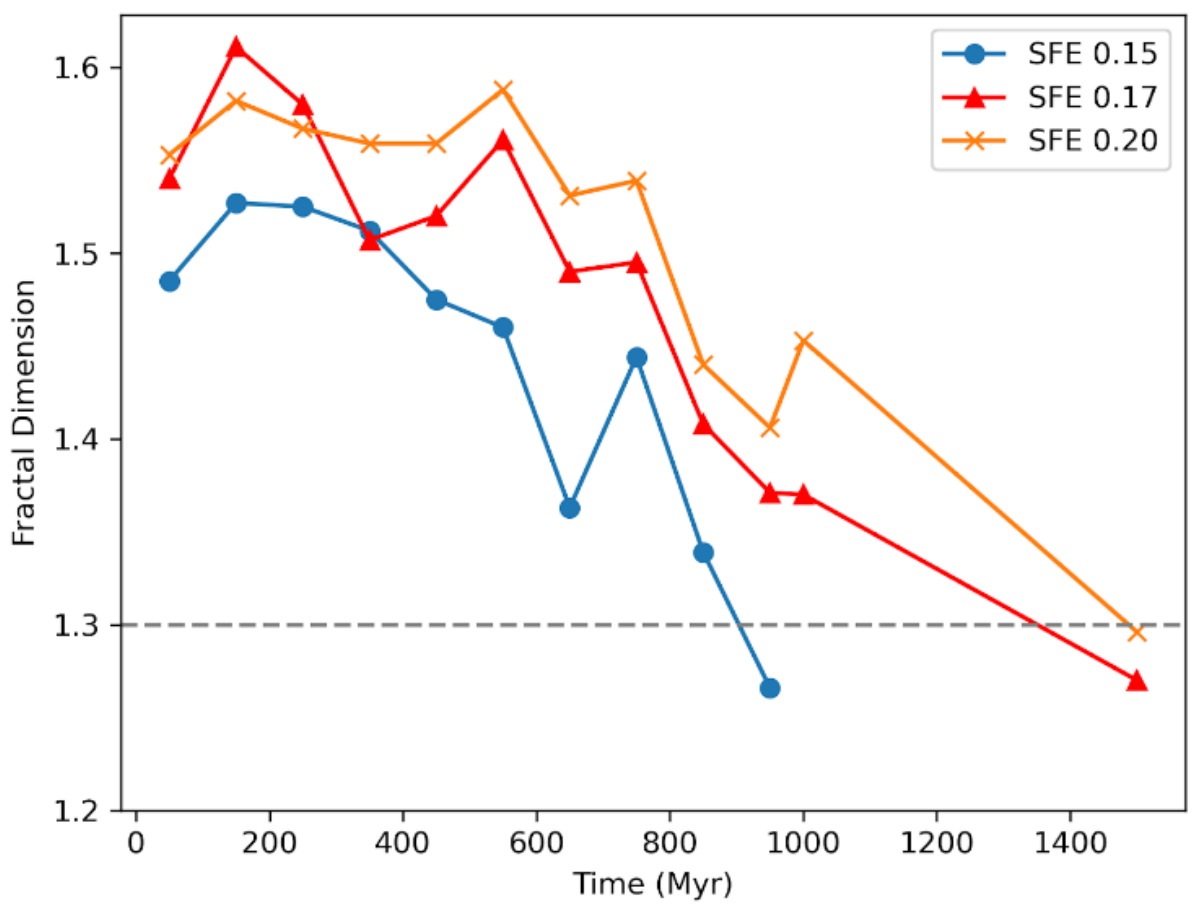

**Fig.3.** Comparison of fractal dimensions of different SFEs according to the evolution of star cluster formation.

To validate our method, we measured the fractal dimensions of model star clusters generated using the publicly available code MCLUSTER [35]. We created star clusters with the following fractal dimensions: 1.6, 2, 2.6, and 3. The results are shown in Figure 4 and Table 1. Due to errors and the inherent randomness of the MCLUSTER model, the fractal dimensions exhibit variations, but the structure in the generated star clusters follows to the power law. Therefore, we can assert that our method effectively captures the fractal structure in the distribution of star clusters.

**Table 1.**

| No. | Fractal dimension with our methods | Fractal dimension with Mcluster |
|---|---|---|
| 1 | 1.35 | 1.6 |
| 2 | 1.44 | 2 |
| 3 | 1.59 | 2.6 |
| 4 | 1.70 | 3 |

.

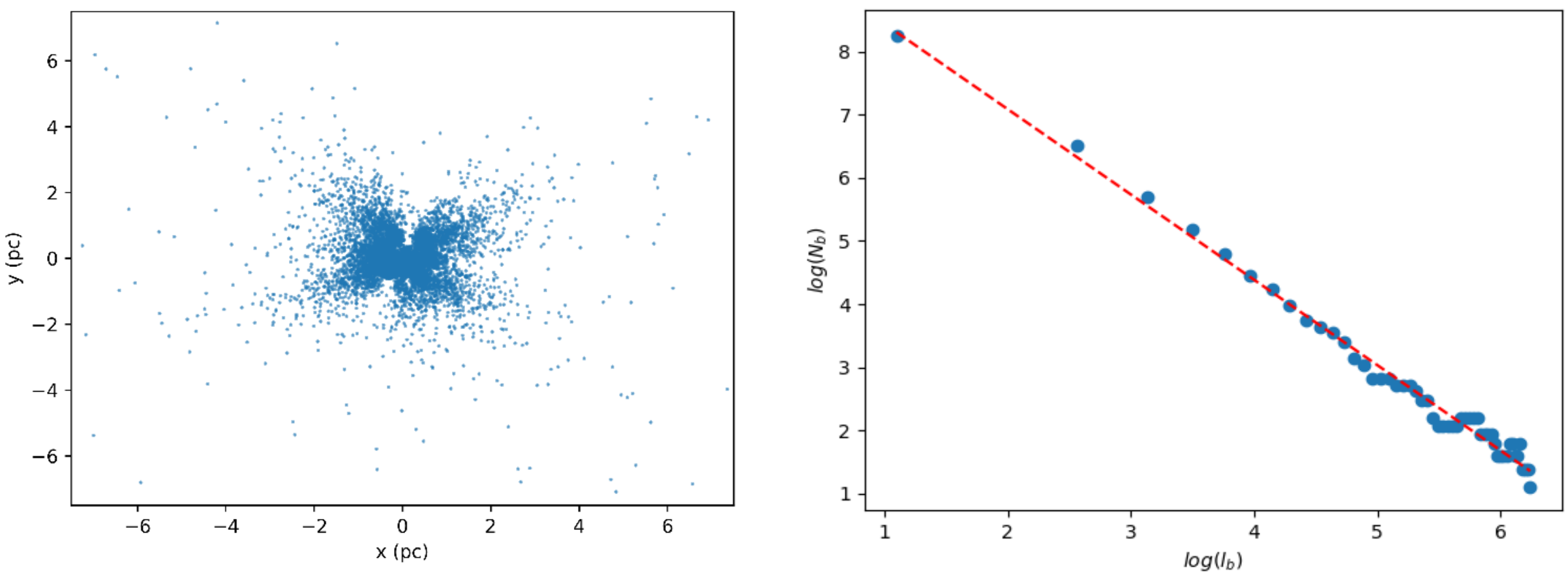

*a)* *b)*

**Fig.4.** a) Distribution of a model star cluster constructed using MCLUSTER for fractal dimension *D = 1.6*; b) Power law for fractal dimension *D = 1.6*.

## 4. Conclusion

In this paper, we used algorithms that had not previously been used in the field of astronomy, specifically in the study of star clusters: MST and BC. To illustrate the proposed method, we modeled star clusters with varying initial SFE from birth to decay. We examined the moments of the cluster's life at intervals of 50 million years and calculated their fractal dimensions within a 1 Jacobi radius. The following conclusions were drawn:

Firstly, a star cluster has a specific fractal dimension and exhibits both fractal and hierarchical structures.

Secondly, the fractal dimension of the star cluster depends on the cluster's structure and varies significantly throughout its evolution. The importance of fractal dimension increases primarily due to rapid dynamic relaxation, which leads to a complexity of the structure. Subsequently, it begins to decrease because of a reduction in the number of stars within the cluster, leading to the structure of the star cluster becoming more homogeneous and less complex.

Thirdly, when the fractal dimension of a star cluster falls below a certain "threshold" (D~1.3), it tends to decay. To compare our results, we used the distribution of star clusters generated by the publicly available code MCLUSTER. Our method was then employed to determine the fractal structure of these clusters.

Our results indicate that our method can serve as an effective tool for studying the structure and topology of complex clusters.

**Conflict of interest statement**

The authors declare that they have no conflict of interest in relation to this research, whether financial, personal, authorship or otherwise, that could affect the research and its results presented in this paper.

**CRediT author statement**

Ussipov N., Akhmetali A.: Methodology, Data curation, Conceptualization; Akniyazova A., Sakan.A.: Writing –original draft, Visualization; Ussipov N., Zaidyn M.: Writing –review & editing, Validation, Supervision; Shukirgaliyev B., Kalambay M.: Formal analysis; Investigation.

The final manuscript was read and approved by all authors.

**Funding**

This research has been funded by the Science Committee of the Ministry of Science and Higher Education of the Republic of Kazakhstan (Grant AP19674715).

**Acknowledgments**

We would like to express our sincerest gratitude to the Department of Solid State Physics and Nonlinear Physics of the Al-Farabi Kazakh National University for supporting this work by providing computing resources (Department of Physics and Technology).

## References

1 Portegies Zwart S. F., McMillan S. L., Gieles M. (2010). Young massive star clusters. *Annual review of astronomy and astrophysics*, 48(1), 431-493. DOI:10.1146/annurev-astro-081309-130834.

2 Lada C.J., Lada E.A. (2003) Embedded clusters in molecular clouds. *Annual Review of Astronomy and Astrophysics*, 41(1), 57-115. DOI: 10.1146/annurev.astro.41.011802.094844.

3 Krumholz M.R., McKee C.F., Bland-Hawthorn J. (2019) Star clusters across cosmic time. *Annual Review of Astronomy and Astrophysics*, 57(1), 227-303. DOI: 10.1146/annurev-astro-091918-104430.

4 Rahner D., Pellegrini E.W., Glover S.C., Klessen R.S. (2019) WARPFIELD 2.0: feedback-regulated minimum star formation efficiencies of giant molecular clouds. *Monthly Notices of the Royal Astronomical Society*, 483(2), 2547-2560. DOI: 10.1093/mnras/sty3295.

5 McLeod A.F., Ali A.A., Chevance M., Della Bruna L., Schruba A., Stevance H.F., Zeidler P. (2021) The impact of pre-supernova feedback and its dependence on environment. *Monthly Notices of the Royal Astronomical Society*, 508(4), 5425-5448. DOI: 10.1093/mnras/stab2726.

6 Banerjee S., Kroupa P. (2013) Did the infant R136 and NGC 3603 clusters undergo residual gas expulsion? *The Astrophysical Journal*, 764(1), 29. DOI: 10.1088/0004-637X/764/1/29.

7 Grasha K., Calzetti D., Adamo A., Kennicutt R.C., Elmegreen B.G., Messa M., Meidt S.E. (2019) The spatial relation between young star clusters and molecular clouds in M51 with LEGUS. *Monthly Notices of the Royal Astronomical Society*, 483(4), 4707-4723. DOI: 10.1093/mnras/sty3424.

8 Kennicutt Jr.R.C. (1989) The star formation law in galactic disks. *Astrophysical Journal,* Part 1, 344, 685-703. DOI: 10.1086/167834.

9 Krause M.G., Offner S.S., Charbonnel C., Gieles M., Klessen R.S., Vázquez-Semadeni E., Zinnecker H. (2020) The physics of star cluster formation and evolution. *Space Science Reviews*, 216, 1 - 46. DOI: 10.1007/s11214-020-00689-4.

10 Murray N. (2011) Star formation efficiencies and lifetimes of giant molecular clouds in the Milky Way. *The Astrophysical Journal*, 729, 2, 133. doi: 10.1088/0004-637X/729/2/133.

11 Higuchi A.E., Kurono Y., Saito M., Kawabe R. (2009) A mapping survey of dense clumps associated with embedded clusters: evolutionary stages of cluster-forming clumps. *The Astrophysical Journal*, *705*(1), 468. doi: 10.1088/0004-637X/705/1/468.

12 Kruijssen J.D., Schruba A., Chevance M., Longmore S.N., Hygate A.P., Haydon D.T., van Dishoeck E.F. (2019) Fast and inefficient star formation due to short-lived molecular clouds and rapid feedback. *Nature*, *569*(7757), 519-522. DOI: 10.1038/s41586-019-1194-3.

13 Lada C.J., Margulis M., Dearborn D. (1984) The formation and early dynamical evolution of bound stellar systems. *Astrophysical Journal,* 285, 141-152. DOI:10.1086/162485.

14 Baumgardt H., Kroupa P. (2007) A comprehensive set of simulations studying the influence of gas expulsion on star cluster evolution. *Monthly Notices of the Royal Astronomical Society*, 380(4), 1589-1598. DOI: 10.1111/j.1365-2966.2007.12209.x.

15 Geyer M.P., Burkert A. (2001) The effect of gas loss on the formation of bound stellar clusters. *Monthly Notices of the Royal Astronomical Society*, 323, 4, 988 – 994. DOI: 10.1046/j.1365-8711.2001.04257.x.

16 Shukirgaliyev B., Parmentier G., Berczik P., Just A. (2017) Impact of a star formation efficiency profile on the evolution of open clusters. *Astronomy & Astrophysics*, 605, A119. DOI: 10.1051/0004-6361/201730607.

17 Shukirgaliyev B., Otebay A., Sobolenko M., Ishchenko M., Borodina O., Panamarev T., Just A. (2021) Bound mass of Dehnen models with a centrally peaked star formation efficiency. *Astronomy & Astrophysics*, 654, A53, DOI: 10.1051/0004-6361/202141299.

18 Lada C.J., Lada E.A. (2003) Embedded clusters in molecular clouds. *Annual Review of Astronomy and Astrophysics*, 41(1), 57-115. DOI:10.1146/annurev.astro.41.011802.094844.

19 Bastian N., Covey K.R., Meyer M.R. (2010) A universal stellar initial mass function? A critical look at variations. *Annual Review of Astronomy and Astrophysics*, 48(1), 339-389. DOI:10.1146/annurev-astro-082708-101642.

20 Grasha K., Calzetti D., Adamo A., Kim H., Elmegreen B.G., Gouliermis D.A., Ubeda L. (2017) The hierarchical distribution of the young stellar clusters in six local star-forming galaxies. *The Astrophysical Journal*, 840(2), 113. DOI: 10.3847/1538-4357/aa6f15.

21 Lahén N., Naab T., Johansson P. H., Elmegreen B., Hu C. Y., Walch S., Moster B.P. (2020) The GRIFFIN Project—Formation of Star Clusters with Individual Massive Stars in a Simulated Dwarf Galaxy Starburst. *The Astrophysical Journal*, 891(1), 2. DOI:10.3847/1538-4357/ab7190.

22 Komjáthy J., Molontay R., Simon K. (2019) Transfinite fractal dimension of trees and hierarchical scale-free graphs. *Journal of Complex Networks*, 7 (5), 764-791. DOI: 10.1093/comnet/cnz005.

23 Mandelbrot B.B. (1983) The fractal geometry of nature. 286. Revised and enlarged edition. New York. DOI:10.1119/1.13295.

24 Kim J.S., Goh K.I., Kahng B., Kim D. (2007) A box-covering algorithm for fractal scaling in scale-free networks. *Chaos: An Interdisciplinary Journal of Nonlinear Science*, 17(2). DOI: 10.1063/1.2737827.

.

25 Wen T., Cheong K.H. (2021) The fractal dimension of complex networks: A review. *Information Fusion*, 73, 87-102. DOI: 10.1016/j.inffus.2021.02.001.

26 Barrow J.D., Bhavsar S.P., Sonoda D.H. (1985) Minimal spanning trees, filaments and galaxy clustering. *Monthly Notices of the Royal Astronomical Society*, 216(1), 17-35. DOI:10.1093/mnras/216.1.17.

27 Bonanno G., Caldarelli G., Lillo F., Mantegna R.N. (2003) Topology of correlation-based minimal spanning trees in real and model markets. *Physical Review E*, 68(4), 046130. DOI: 10.1103/ PhysRevE.68.046130.

28 Bhavsar S.P., Ling E.N. (1988) II. Large-Scale Distribution of Galaxies: Filamentary Structure and Visual Bias. *Publications of the Astronomical Society of the Pacific*, 100(633), 1314. DOI:10.1086/132325.

29 Libeskind N.I., Van De Weygaert R., Cautun M., Falck B., Tempel E., Abel T., Yepes G. (2018) Tracing the cosmic web. *Monthly Notices of the Royal Astronomical Society*, 473(1), 1195-1217. DOI:10.48550/arXiv.1705.03021 .

30 Allison R.J., Goodwin S.P., Parker R.J., De Grijs R., Zwart S.F.P., Kouwenhoven M.B.N. (2009) Dynamical mass segregation on a very short timescale. *The Astrophysical Journal*, *700*(2), L99. DOI:10.1088/0004-637X/700/2/L99.

31 Naidoo K., Whiteway L., Massara E., Gualdi D., Lahav O., Viel M., Font-Ribera A. (2020). Beyond two-point statistics: using the minimum spanning tree as a tool for cosmology. *Monthly Notices of the Royal Astronomical Society*, 491(2), 1709-1726. DOI:10.1093/mnras/stz3075.

32 Plummer H.C. (1911) On the problem of distribution in globular star clusters. *Monthly Notices of the Royal Astronomical Society,* 71, 460-470. DOI: 10.1093/mnras/71.5.460.

33 Zhanabaev Z., Ussipov N., Khokhlov S. (2021) Scale-invariant and wave nature of the Hubble parameter. *Eurasian Physical Technical Journal*, 18(2 (36)), 81-89. DOI:10.13140/RG.2.2.31792.61449/1.

34 Küpper A.H., Maschberger T., Kroupa P., Baumgardt H. (2011) Mass segregation and fractal substructure in young massive clusters–I. The McLuster code and method calibration. *Monthly Notices of the Royal Astronomical Society*, 417(3), 2300-2317. DOI: 10.1111/j.1365-2966.2011.19412.x.

---

## AUTHORS' INFORMATION

**Ussipov, Nurzhan –** Senior Lecturer, Department of Physics and Technology, Al-Farabi Kazakh National University, Almaty, Kazakhstan; Scopus Author ID: 57226319348; ORCID: 0000-0002-2512-3280; ussipov.nurzhan @kaznu.kz,

**Akhmetali, Almat** – Bachelor (Sci.), Department of Physics and Technology, Al-Farabi Kazakh National University, Almaty, Kazakhstan; Scopus Author ID: 58759186800; ORCID: 0009-0005-7254-524X; akhmetalialmat @gmail.com

**Zaidyn, Marat** – Student, Department of Physics and Technology, Al-Farabi Kazakh National University, Almaty, Kazakhstan, Scopus Author ID: 59194267200; ORCID: 0009-0006-8505-7277; zaidyn_marat@live.kaznu.kz

**Akniyazova, Aigerim** – Master (Sci.), Lecturer, Department of Physics and Technology, Al-Farabi Kazakh National University, Almaty, Kazakhstan; ORCID: 0000-0002-9185-3185; aigerimakniyazova@gmail.com

**Sakan, Aknur** – Bachelor of natural sciences, Department of Physics and Technology, Al-Farabi Kazakh National University, Almaty, Kazakhstan; ORCID: 0009-0001-8784-4470 ; aknursakan47@gmail.com

**Mukhagali, Kalambay** – PhD, Senior Lecturer, Heriot-Watt branch in Kazakhstan, K.Zhubanov Aktobe Regional University, Aktobe, Kazakhstan; Scopus Author ID: 57224666055; ORCID:0000-0002-0570-7270; mukhagaly@gmail.com

**Bekdaulet, Shukirgaliyev –** PhD, Associate Professor, Heriot-Watt branch in Kazakhstan, K.Zhubanov Aktobe Regional University, Aktobe, Kazakhstan; Scopus Author ID: 57163629900; ORCID: 0000-0002-4601-7065; b.shukirgaliyev@hw.ac.uk

## APPENDIX 1

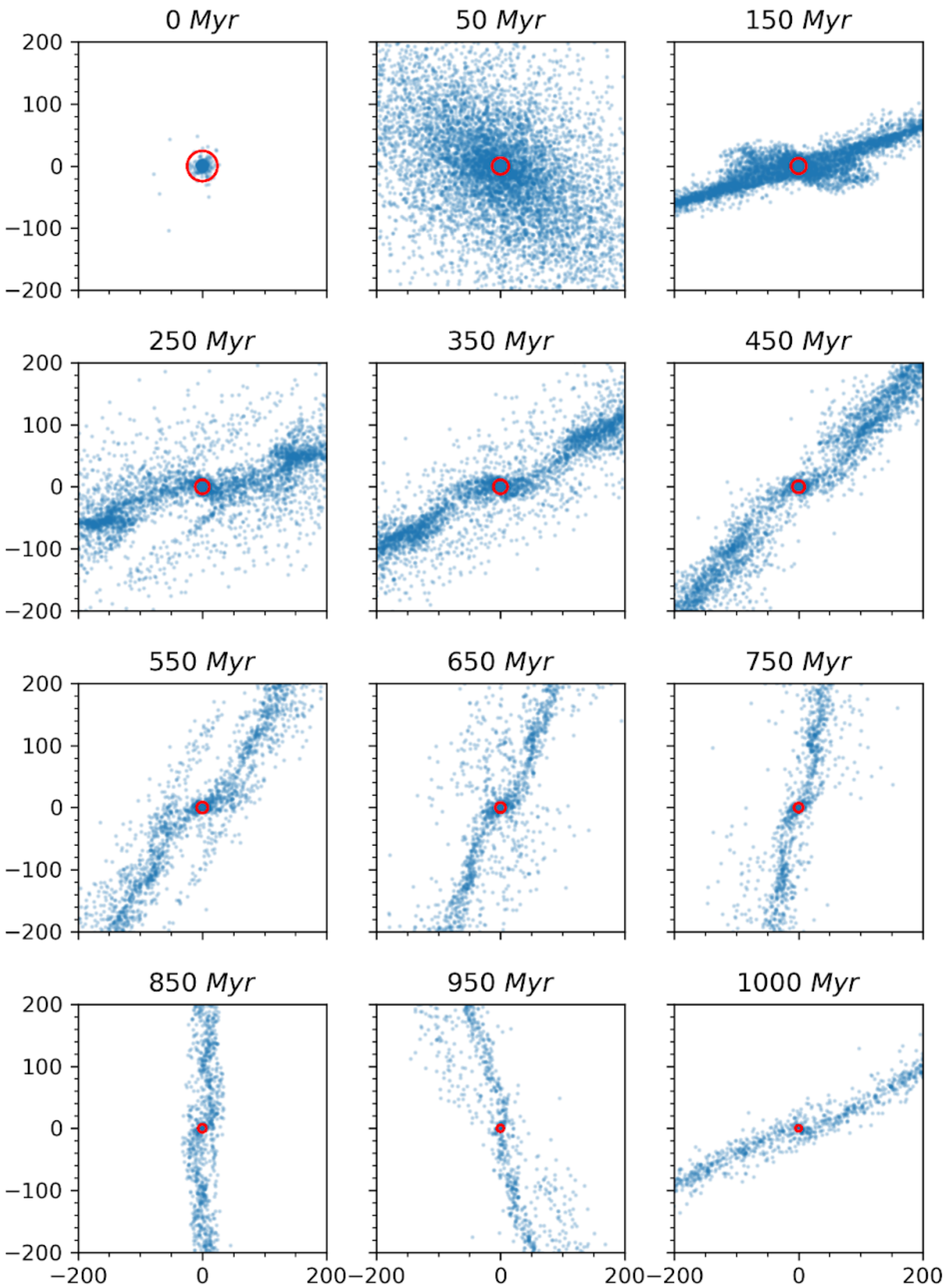

**Fig.A1.** Evolution of the cluster with SFE=0.15 (red circle is a Jacobi radius).

Figure A1 shows the evolution of the cluster with an SFE=15% and a time interval of 50 Myr. The red circle marks the Jacobi radius which decreases as the cluster dissolves.

.

## APPENDIX 2

**Table A1.** Statistical analysis of clusters: SFE=0.15, 0.17, 0.20. In Table A1, the Jacobi radius at various stages of SC's lifetime, along with mass and number of stars within that radius, are shown.

| Age (Myr) | SFE=0.15 | | | SFE=0.17 | | | SFE=0.20 | | |
|---|---|---|---|---|---|---|---|---|---|
| | $R_J$ (pc) | N | $M_J$ ($M_\odot$) | $R_J$ (pc) | N | $M_J$ ($M_\odot$) | $R_J$ (pc) | N | $M_J$ ($M_\odot$) |
| **0** | 24.40 | 10454 | 6000 | 24.44 | 10433 | 6000 | 24.51 | 10432 | 6000 |
| **50** | 13.50 | 2138 | 998.1 | 15.19 | 3068 | 1424.8 | 16.33 | 3902 | 1766.3 |
| **150** | 12.50 | 1803 | 793.7 | 14.44 | 2768 | 1224.0 | 15.53 | 3502 | 1521.6 |
| **250** | 11.35 | 1307 | 591.9 | 13.51 | 2287 | 1007.4 | 14.70 | 3018 | 1291.3 |
| **350** | 10.55 | 1002 | 476.4 | 13.03 | 1958 | 898.8 | 14.21 | 2675 | 1161.5 |
| **450** | 9.80 | 750 | 383.2 | 12.53 | 1661 | 799.9 | 13.65 | 2304 | 1034.2 |
| **550** | 8.88 | 532 | 283.9 | 11.94 | 1376 | 691.6 | 13.19 | 2009 | 931.5 |
| **650** | 8.14 | 356 | 219.4 | 10.40 | 1140 | 601.4 | 12.71 | 1712 | 833.1 |
| **750** | 7.21 | 211 | 152.5 | 10.74 | 899 | 503.07 | 12.08 | 1390 | 718.7 |
| **850** | 6.52 | 137 | 112.3 | 10.00 | 673 | 406.8 | 11.58 | 1150 | 632.2 |
| **950** | 5.47 | 75 | 66.6 | 9.11 | 454 | 307.5 | 11.01 | 931 | 542.9 |
| **1000** | 4.90 | 51 | 47.8 | 8.72 | 377 | 269.7 | 10.83 | 840 | 514.6 |